\documentclass[12pt,twoside]{article}
\usepackage{amssymb}
\usepackage{amsmath}
\usepackage{mathabx}
\usepackage{latexsym}
\usepackage{longtable}
\usepackage{graphicx,bbm,psfrag}
\usepackage{cite}

\begin{document}
\begin{titlepage}
\vspace*{5cm}
\begin{center}
\Large{\textbf{Currents of the Ablowitz-Ladik Chain 
}}\bigskip\bigskip\bigskip
\end{center}
\begin{center} 
{\large{Herbert Spohn}}\bigskip\bigskip\\
Departments of Mathematics and Physics, Technical University Munich,\smallskip\\
Boltzmannstr. 3, 85747 Garching, Germany
\end{center}
\vspace{3cm}
\begin{flushright}
May 1, 2024
\end{flushright}
\vspace{2cm}
\textbf{Abstract}: While the currents of the defocusing Ablowitz-Ladik chain are unambiguously defined, no explicit formulas for their densities have been obtained yet. 
We close this gap.  As a consequence, we also complete the argument which establishes the anticipated expression for average currents, averaged over a generalized Gibbs ensemble. 
\end{titlepage}
\section{Introduction, defocusing Ablowitz-Ladik chain}
\label{sec1}
\setcounter{equation}{0}
The Ablowitz-Ladik chain is an integrable discretization of the nonlinear Schr\"{o}dinger equation and governs the dynamics of  
a complex-valued wave field $\{\alpha_j \in \mathbb{C}\}$ with  $j \in \mathbb{Z}$ or possibly some finite interval, 
$j \in [1,\ldots,N]$. For the defocusing case one requires $|\alpha_j| \leq 1$. Then the equations of motion are    
\begin{equation}\label{1.1} 
\frac{d}{dt}\alpha_j  = \mathrm{i} \rho_j^2 (\alpha_{j-1} + \alpha_{j+1}),\quad \rho_j^2 = 1 - |\alpha_j|^2.
\end{equation}
The inverse scattering transform for the AL chain has been developed early on \cite{AL75,AL76}.   A most readable account is the book \cite{APT04}
which covers a wide range of related material.
In a very recent contribution, we report on an underlying Hamiltonian particle dynamics for which 
particles retain their order. In such framework  the asymptotic $N$-particle relative scattering shifts are given by a sum of two-particle shifts  \cite{BS24}.

For the purpose of this study, we rely on the Lax matrix, first constructed by Nenciu \cite{N05}, see also \cite{N06,KN07}. For the \textit{closed}
chain one imposes periodic boundary conditions, $\alpha_{N+1} = \alpha_1$, and restricts to even $N$  for convenience.
Before actually turning to the Lax matrix, one notes that
the evolution equations \eqref{1.1} are of Hamiltonian form by 
regarding $\alpha_j$ and its complex conjugate $\bar{\alpha}_j$ as canonically conjugate variables and introducing
the weighted Poisson bracket
\begin{equation}\label{1.2} 
\{f,g\}_\mathrm{AL} =\mathrm{i} \sum_{j=1}^{N} \rho_j^2\Big(\frac{\partial f}{\partial{\bar{\alpha}_j}} 
\frac{\partial g}{\partial{\alpha_j}}   - \frac{\partial f}{\partial{\alpha_j}} 
\frac{\partial g}{\partial{\bar{\alpha}_j}} 
\Big).
\end{equation}
The Hamiltonian of the AL system reads
\begin{equation}\label{1.3} 
H_N = - \sum_{j=1}^{N} \big( \alpha_{j-1} \bar{\alpha}_{j} + \bar{\alpha}_{j-1} \alpha_{j}\big).
\end{equation}
One readily checks that indeed
\begin{equation} \label{1.4}
\frac{d}{dt}\alpha_j = \{\alpha_j,H_N\}_\mathrm{AL}   = \mathrm{i} \rho_j^2 (\alpha_{j-1} + \alpha_{j+1}) .
\end{equation}

The Lax matrix has the structure of a   
Cantero-Moral-Vel\'{a}zquez (CMV) matrix \cite{CMV03,CMV05,KN07,S07}. The basic building blocks are the $2\times 2$ matrices,
\begin{equation}\label{1.5} 
\Xi_j =
\begin{pmatrix}
\bar{\alpha}_j & \rho_j \\
\rho_j & -\alpha_j  \\
\end{pmatrix},
\end{equation}
and one constructs two $N\times N$ matrices as
\begin{equation}\label{1.6} 
 L_N = \mathrm{diag}(\Xi_1,\Xi_3,\dots, \Xi_{N-1}), \quad \big( M_N\big)_{i,j = 2,...,N-1} = \mathrm{diag}(\Xi_2,\Xi_4,\dots, \Xi_{N-2}),
\end{equation}
together with $(M_N)_{1,1} =  -\alpha_{N}$, $(M_N)_{1,N} = \rho_{N}$, $(M_N)_{N,1} =  \rho_{N}$, and $(M_N)_{N,N-} = \bar{\alpha}_{N}$. More pictorially, $L_N$ corresponds to the blocks $(1,2),...,(N-1,N)$ while $M_N$ uses the by $1$ shifted blocks $(2,3),...,(N,1)$. The CMV matrix associated to the coefficients $\alpha_1,\dots,\alpha_{N}$ is then given by 
\begin{equation} \label{1.7}
C_N = L_N M_N.
\end{equation}
$C_N$ is the Lax matrix of the AL chain. Obviously, $L_N,M_N$ are unitary and so is $C_N$. The  eigenvalues of $C_N$ are denoted by $\mathrm{e}^{\mathrm{i}\vartheta_j}$, $\vartheta_j \in [0,2 \pi]$,
$j = 1,\dots,N$.  Of course, the eigenvalues depend on $N$ which is suppressed in our notation.

Next we define for a general matrix, $A$, the $+$ operation as
\begin{equation} \label{1.8}
 (A_+)_{i,j} = \begin{cases} A_{i,j}  &\mathrm{if} \,\,i < j,\\
   \tfrac{1}{2}A_{i,j} &\mathrm{if} \,\,i = j,\\
  0 &\mathrm{if} \,\,i > j.
 \end{cases}
\end{equation}
Then one version of the Lax pair reads
\begin{equation}\label{1.9}
\{C_N,\mathrm{tr}[C_N]\}_\mathrm{AL} = \mathrm{i} [C_N,C_{N+}],\qquad \{C_N,\mathrm{tr}[C_N^*]\}_\mathrm{AL} = \mathrm{i} [C_N,(C_{N+})^*],
\end{equation}
where $A^*$ denotes the hermitian adjoint of $A$. Since the Poisson bracket acts as a derivative, one deduces
\begin{equation} \label{1.10}
\{(C_N)^n,\mathrm{tr}[C_N]\}_\mathrm{AL} = \sum_{m=0}^{n-1} (C_N)^m\mathrm{i}[C_N,C_{N+}](C_N)^{n-m-1} = \mathrm{i} [(C_N)^n,C_{N+}], 
\end{equation}
and similarly 
\begin{equation} \label{1.11}
\{(C_N)^n,\mathrm{tr}[C_N^*]\}_\mathrm{AL} =  \mathrm{i} [(C_N)^n,(C_{N+})^*]. 
\end{equation}
Hence the locally conserved fields are given by
\begin{equation} \label{1.12}
Q^{[n],N} = \mathrm{tr}\big[(C_N)^n\big].
\end{equation}
By a similar argument, it can be shown that the mutual Poisson brackets vanish,
\begin{equation} \label{1.13}
\{Q^{[n],N}, Q^{[n'],N}\}_\mathrm{AL} = 0 \bigskip.
\end{equation}
In addition, conserved is the log intensity field
\begin{equation}\label{1.14}
Q^{[0],N} =  - \sum_{j=1}^{N}  \log(\rho_j^2).
\end{equation}

Our goal are the currents averaged with respect to a generalized Gibbs ensemble (GGE). 
The natural a priori measure on the ring is the product measure
\begin{equation} \label{1.15}
\prod_{j=1}^{N}\mathrm{d}^2\alpha_j (\rho_j^2)^{P-1}= \prod_{j=1}^{N}\mathrm{d}^2\alpha_j (\rho_j^2)^{-1}\exp\big(-PQ^{[0],N}\big) 
\end{equation}
on $\mathbb{D}^N$, $\mathbb{D} = \{z||z|\leq 1\}$. This measure is invariant under the dynamics generated by \eqref{1.1}. To be able to normalize the measure, $P>0$ is  required.  The log intensity is controlled by the parameter $P$ which, in analogy to the 
Toda lattice, is called pressure. Small $P$ corresponds to maximal log intensity, i.e. $|\alpha_j|^2 \to 1$,  and large $P$ to low log intensity, i.e. $|\alpha_j|^2 \to 0$. 
In the grand-canonical ensemble, the Boltzmann weight is constructed as the exponential of an arbitrary linear combination of the conserved fields. More concisely, and in greater generality this ensemble can written as
\begin{equation} \label{1.16}
Z_{N}(P,V)^{-1}\prod_{j=1}^{N}\mathrm{d}^2\alpha_j (\rho_j^2)^{P-1}\exp\big(-\mathrm{tr}[V(C_N)]\big), \qquad P >0 .
\end{equation}
$Z_{N}(P,V)$ is the normalizing partition function. The confining potential $V$ is a real-valued continuous function on the unit circle, independent of $N$.

The generalized free energy per unit length is defined by 
\begin{equation} \label{1.17}
F(P,V) = -\lim_{N \to\infty}\frac{1}{N}\log Z_{N}(P,V).
\end{equation}
Physically more relevant is the empirical density of states (DOS) given by 
 \begin{equation} \label{1.18} 
 \varrho_N(z) = \frac{1}{N} \sum_{j=1}^N \delta(\mathrm{e}^{\mathrm{i}\vartheta_j} -z),
 \end{equation}
 as a probability measure on the complex unit circle. The DOS is random under the GGE and one would like to understand its large $N$ limit. This problem has  recently been studied from the probabilistic side  \cite{MG21,MM22,M23}.
 Proved is the existence  of the infinite volume Gibbs measure, its uniqueness, and  exponential mixing. For the DOS established  is the almost sure convergence to a limiting 
probability  measure on the unit circle, which has a density relative to $\mathrm{d}\vartheta$ and is characterized as the unique minimum of a variational principle.
%%%%%%%%%%%%%%%%%%%%%%%%%%%%%%%%%%%%%%%%%%%%
%%%%%%%%%%%%%%%%%%%%%%%%%%%%%%%%%%%%%%%%%%%%
\section{The GGE averaged currents}
\label{sec2}
\setcounter{equation}{0}
We start from the conserved field $\mathrm{tr}[(C_N)^n]$ which is complex-valued. The field has the density $Q_j^{[n]}= (C^{n})_{j,j}$, where $C^{n}$ is the $n$-th power of the infinite volume matrix, $C$, defined as an operator acting in $\ell_2(\mathbb{Z})$.
The matrix elements $(C^{n})_{j,j}$ depend only on $\{\alpha_i|i \in [j-2n -1,j+2n+1]\}$. Hence we can define the complex currents, $ J_{j}^{[n]}$, implicitly through
\begin{equation}
\label{2.1} 
\frac{\mathrm{d}}{\mathrm{d}t} Q_j^{[n]} = \{Q_j^{[n]},H_{\mathrm{al}}\}_\mathrm{AL} = J_{j}^{[n]} - J_{j+1}^{[n]},
\end{equation}
where 
\begin{equation}
\label{2.1a} 
H_{\mathrm{al}} = \mathrm{tr}[C + C^*].
\end{equation}
Because of the conservation law, such current densities exist and are again supported on a finite interval,
see e.g. the discussion in \cite{S24}. So far missing has been a more explicit formula, which is required for computing the GGE averaged currents.
The Lax matrix is only invariant under shifts by $2$.  A convenient starting formula is therefore the sum
\begin{equation}
\label{2.2} 
\{(C^{n})_{0,0} + (C^{n})_{1,1},H_{\mathrm{al}}\}_\mathrm{AL} = J_{0}^{[n]} - J_{2}^{[n]}.
\end{equation}

Using Eq. \eqref{1.9} together with the fact that the matrix $C_+$ has non-vanishing matrix elements only for
$(j,j+\ell)$ with $\ell = 0,1,2$, the Poisson bracket  is obtained to
\begin{eqnarray} \label{2.3}
&&\hspace{-40pt}\{(C^n)_{j,j},H_\mathrm{al}\}_\mathrm{AL} \nonumber\\
&&\hspace{-30pt} = \sum_{\ell = 1,2}\mathrm{i}\Big((C^n)_{j,j-\ell} C_{j-\ell,j} -C_{j,j+\ell}(C^n)_{j+\ell,j} + (C^n)_{j,j+\ell} \bar{C}_{j,j+\ell} - \bar{C}_{j-\ell,j}(C^n)_{j-\ell,j}\Big).  
\end{eqnarray}
The term with $\ell = 0$ does not contribute. The matrix element $(0,0)$ in \eqref{2.3} reads
\begin{eqnarray} \label{2.4}
&&\hspace{-50pt}\{(C^n)_{0,0},H_\mathrm{al}\}_\mathrm{AL} = \mathrm{i}\big((C^n)_{0,-1}C_{-1,0} +(C^n)_{0,-2} C_{-2,0} - C_{0,1}(C^n)_{1,0} - C_{0,2}(C^n)_{2,0}\nonumber\\[1ex]
&&\hspace{46pt}+(C^n)_{0,1}\bar{C}_{0,1} +(C^n)_{0,2} \bar{C}_{0,2} -\bar{C}_{-1,0}(C^n)_{-1,0} -\bar{C}_{-2,0}(C^n)_{-2,0}\big)
\end{eqnarray}
and the matrix element $(1,1)$ equals 
\begin{eqnarray} \label{2.5}
&&\hspace{-50pt}\{(C^n)_{1,1},H_\mathrm{al}\}_\mathrm{AL} = \mathrm{i}\big((C^n)_{1,0}C_{0,1} + (C^n)_{1,-1}C_{-1,1} - C_{1,2}(C^n)_{2,1} - C_{1,3}(C^n)_{3,1}\nonumber\\[1ex]
&&\hspace{46pt}+(C^n)_{1,2}\bar{C}_{1,2} + (C^n)_{1,3}\bar{C}_{1,3} -\bar{C}_{0,1}(C^n)_{0,1} -\bar{C}_{-1,1}(C^n)_{-1,1}\big).
\end{eqnarray}
Up to a shift by $2$, according to \eqref{1.7}, the matrix elements for even left index are 
\begin{equation}\label{2.6}
C_{0,-1} = \rho_{-1} \bar{\alpha}_0, \quad C_{0,0} = - \alpha_{-1} \bar{\alpha}_0, \quad C_{0,1} = \rho_{0} \bar{\alpha}_1, \quad C_{0,2} = \rho_0 \rho_1, 
\end{equation}
and for odd left index
\begin{equation}\label{2.7}
C_{1,-1} = \rho_{-1} \rho_0, \quad C_{1,0} = - \alpha_{-1} \rho_0, \quad C_{1,1} = -  \alpha_{0} \bar{\alpha}_1, \quad C_{1,2} = -\alpha_0 \rho_1.
 \end{equation}
All other matrix elements vanish. Inserting in \eqref{2.5}, one notes that  $C_{-1,1} = 0$ and $C_{1,3} = 0$.  Two terms in \eqref{2.4} cancel against the respective two terms in \eqref{2.5}. The remaining terms 
are paired as a difference of distance two. According to definition \eqref{2.2}, this yields for the current 
\begin{equation}\label{2.8}
J_0^{[n]} = \mathrm{i}\big(\rho_{-2} \rho_{-1}(C^n)_{0,-2} - \rho_{-2} \rho_{-1}(C^n)_{-2,0}  -\alpha_{-2}  \rho_{-1} (C^n)_{0,-1} + \bar{\alpha}_{-2} \rho_{-1}(C^n)_{-1,0}\big).
 \end{equation}
 
 To compute the GGE averaged currents, one  notes that 
\begin{equation}\label{2.9}
J^{[0]}_j = - \mathrm{i}(Q_j^{[1]} - \bar{Q}_j^{[1]}). 
\end{equation}
With this simple identity one can use the symmetry of the charge-current correlator to express the GGE averaged current 
through the fluctuation covariance  of the DOS. The detailed argument is provided in \cite{S24} and leads to 
\begin{equation}\label{2.10}
\partial_P\big(\langle J^{[n]}_0 \rangle_{P,V} + 2P\langle \varsigma_{1-}, C^\sharp \varsigma_{n}\rangle\big) = 0.
\end{equation}
The expression in the round brackets is independent of $P$. But one would like to show that it actually vanishes.
Since $\langle \varsigma_{1-}, C^\sharp \varsigma_{n}\rangle$ is bounded in $P$, the second summand vanishes at $P=0$.
Thus we are left with establishing $\langle J^{[n]}_0 \rangle_{P=0,V} =0$.

 In the limit $P \to 0$, for each $j$ the a priori measure \eqref{1.15} concentrates to the  uniform  measure on the unit circle. In particularly, $\langle \rho_j^2\rangle_P \to 0$  as $P\to 0$ and the  Lax matrix turns diagonal. Denoting 
$\alpha_j = \mathrm{e}^{\mathrm{i}\phi_j}$, $\phi_j \in [0,2\pi]$, in the limit $P \to 0$ the GGE \eqref{1.17}  converges  to
\begin{equation} \label{2.11}
\frac{1}{Z_{\mathrm{al},N}(0,V)}\prod_{j=1}^{N}\mathrm{d}\phi_j \exp\Big( - \sum_{j=1}^{N} V\big(- \mathrm{e}^{\mathrm{i}(\phi_{j} - \phi_{j+1})}\big)\Big)
\end{equation}
with boundary conditions $\phi_{N+1} = \phi_1$. Now in each term of the expression \eqref{2.8} for the current there is a factor of $\rho_j$. 
As $P \to 0$, also $\rho_j \to 0$ in probability. Hence 
 \begin{equation}\label{2.12} 
 \lim_{P \to 0} \langle J^{[n]}_0 \rangle_{P,V} =0. 
 \end{equation}
In \cite{S21} the GGE averaged currents  and the equations of GHD for the AL chain were written simply assuming the validity of the limit \eqref{2.12}.
%%%%%%%%%%%%%%%%%%%%%%%%%%%%%%%%%%%%%%%%%%%%%
%%%%%%%%%%%%%%%%%%%%%%%%%%%%%%%%%%%%%%%%%%%
\section{Schur flow and its averaged currents }
\label{sec3}
\setcounter{equation}{0}
The Schur flow refers to the Hamiltonian
\begin{equation}\label{3.1} 
H_{\mathrm{kv},N} = \sum_{j=1}^{N} (- \mathrm{i})\big( \alpha_{j-1} \bar{\alpha}_{j} - \bar{\alpha}_{j-1} \alpha_{j}\big) = - \mathrm{i}\,\mathrm{tr}[C_N - C_N^*].
\end{equation}
The index kv should be a reminder that the formal continuum limit of the dynamics generated by \eqref{3.1} is the modified Korteweg-de Vries equation.  
The Hamiltonians $H_{\mathrm{kv},N}$ and  $H_{\mathrm{al},N}$ differ only by signs and factors of $\mathrm{i}$. Therefore we can record immediately the starting formula
\begin{eqnarray} \label{3.2}
&&\hspace{-40pt}\{(C^n)_{j,j},H_\mathrm{kv}\}_\mathrm{AL} \nonumber\\
&&\hspace{-30pt} = \sum_{\ell = 1,2}\Big((C^n)_{j,j-\ell} C_{j-\ell,j} -C_{j,j+\ell}(C^n)_{j+\ell,j} - (C^n)_{j,j+\ell} \bar{C}_{j,j+\ell} + \bar{C}_{j-\ell,j}(C^n)_{j-\ell,j}\Big).  
\end{eqnarray}
The GGE averaged current is then 
\begin{equation}\label{3.3}
J_{\mathrm{kv},0}^{[n]} =  \rho_{-2} \rho_{-1}(C^n)_{0,-2} +  \rho_{-2} \rho_{-1}(C^n)_{-2,0} -  \alpha_{-2} \rho_{-1}\ (C^n)_{0,-1} - \bar{\alpha}_{-2}\rho_{-1}
(C^n)_{-1,0}.
 \end{equation}
 The stretch current turns into
 \begin{equation}\label{3.4}
J^{[0]}_j = Q_j^{[1]} + \bar{Q}_j^{[1]}. 
\end{equation}
As above, one still concludes that $\langle J^{[n]}_0 \rangle_{P=0,V} =0$. 

The interesting case is however based on the observation that under the Schur flow, real initial data stay real. 
Then $\alpha_j$ becomes a real-valued field and the integration in the a priori measure \eqref{1.15} is over the hypercube $[-1,1]^N$. The stretch current equals 
$2J^{[0]}_j = Q_j^{[1]}$ and one can still use the symmetry of the charge current correlator.
In the limit $P\to 0$ the a priori measure converges to 
\begin{equation} \label{3.5}
\prod_{j=1}^{N}\mathrm{d}\alpha_j \big(\delta(\alpha_j -1) + \delta(\alpha_j +1)\big).
\end{equation}
In particularly $\rho_j \to 0$ and, since every term in \eqref{3.3} has a prefactor of $\rho_j$, one concludes that the analogue of \eqref{2.12} still holds.

In conclusion, the GGE averaged currents have been determined for the AL chain with two distinct Hamiltonians, both linear in $C$ and $C^*$. Thereby we added one 
further model for which the collision rate ansatz is proved by relying on the symmetry of the charge-current correlator  \cite{S20}, \cite{YS20}.

\end{document}